\documentclass[onecolumn,showpacs,preprintnumbers,amsmath,amssymb,showkeys,12ft]{revtex4}

\usepackage{graphicx}
\usepackage{dcolumn}
\usepackage{bm}

\begin{document}
\draft
\date{\today}
\title{On The Problem of Constraints In Nonextensive Formalism: A Quantum Mechanical Treatment}
\author{G. B. Ba\u{g}c\i $^{a}$, Altu\u{g} Arda$^{b}$, Ramazan Sever$^{c}$
}\thanks{Corresponding Author\\}
\email{sever@newton.physics.metu.edu.tr}
\address{$^{a}$ Department of Physics, University of North Texas, P.O. Box 311427, Denton, TX 76203-1427,
USA\\
 $^{b}$ Physics Education, Hacettepe University, 06532,\\
Ankara, Turkey\\
$^{c}$ Department of Physics, Middle East Technical University, 06531,\\
Ankara, Turkey}

\pagenumbering{arabic}

\begin{abstract}
Relative entropy (divergence) of Bregman type recently proposed by
T. D. Frank and Jan Naudts is considered and its quantum counterpart
is used to calculate purity of the Werner state in nonextensive
formalism. It has been observed that Bregman type divergence is
suitable for q
\mbox{$>$}%
1 whereas Csisz\`{a}r type is suitable for q$\in (0,1)$. It is then
argued that the difference is due to the fact that the relative
entropy of Bregman type is related to the first choice
thermostatistics whereas the one of Csisz\`{a}r type is related to
the third choice thermostatistics. Moreover, it has been noted that
these two measures show different qualitative behavior with respect
to F due to the fact that these divergences being written compatible
with different constraints. The possibility of writing a relative
entropy of Bregman type compatible with the third choice has been
investigated further. The answer turns out to be negative as far as
the usual transformation from ordinary probabilities to the escort
probabilities are considered.
\end{abstract}

\pacs{PACS: 05.20.-y; 05.30.-d; 05.70. ; 03.65.-w}  \narrowtext
\newpage \setcounter{page}{1}
\keywords{quantum divergence,nonextensivity,escort probability}
\maketitle

\section{\protect\bigskip Introduction}

\noindent \qquad A nonextensive generalization of the standard
Boltzmann-Gibbs (BG) entropy has been proposed by C. Tsallis in 1988
[1-4]. This new definition of entropy is given by

\begin{equation}
S_{q}=k\frac{1-\sum_{i=1}^{W}p_{i}^{q}}{q-1},
\end{equation}
where k is a positive constant which becomes the usual Boltzmann
constant in the limit q$\rightarrow 1$, p$_{\text{i }}$\bigskip is
the probability of the system in the ith microstate, W is \ the
total number of the configurations of the system.The entropic index
q is a real number, which characterizes the degree of nonextensivity
as can be seen from the following pseudo-additivity rule:

\begin{equation}
S_{q}(A+B)/k=[S_{q}(A)/k]+[S_{q}(B)/k]+(1-q)[S_{q}(A)/k][S_{q}(B)/k],
\end{equation}
where A and B are two independent systems i.e., p$_{ij}$(A+B)$=$p$_{i}(A)$p$%
_{j}(B)$. As q$\rightarrow 1$, the nonextensive entropy definition
in Eq. (1) becomes

\begin{equation}
S_{q\rightarrow 1}=-k_{B}\sum_{i=1}^{W}p_{i}\ln p_{i},
\end{equation}
which is the usual BG entropy. This means that the definition of
nonextensive entropy contains BG statistics as a special case. The
cases q
\mbox{$<$}%
1, q
\mbox{$>$}%
1 and $q=1$ correspond to superextensivity, subextensivity and
extensivity, respectively.

The nonextensive formalism is used in systems with long-ranged
interactions, long-ranged memories, and systems which evolve in
fractal-like space-time. Even though BG statistics can be used
successfully in investigating extensive systems, physical systems
such as Euler two-dimensional turbulence [5], high energy collisions
[6-9], nematic liquid crystals [10], stellar polytropes [11], the
nonisotropic rigid rotator model [12] and Fokker-Planck systems
[13,14] can be given as examples for which the nonextensive
formalism has been used successfully.

\bigskip From a mathematical point of view, the nonextensive formalism is
being formulated by using the q-deformed logarithm and q-deformed
exponential which can be given, respectively, as

\begin{equation}
\ln _{q}x\equiv \frac{x^{1-q}-1}{1-q}\text{ \ \ \ \ \ \ \ \ \ \ \ \
\ \ \ \ \ exp}_{q}x\equiv \lbrack 1+(1-q)x]^{1/(1-q)}.
\end{equation}

These functions become the usual logarithmic and exponential
functions as q$\rightarrow 1.$ Moreover, other functions in common
use can be generalized in a similar manner. For example q-sine
function [15] can be defined as

\begin{equation}
\mathop{\rm Si}%
n_{q}x\equiv \sum_{j=0}^{\infty
}\frac{(-1)^{j}Q_{2j}x^{2j+1}}{(2j+1)!},
\end{equation}
where the function Q$_{n}(q)$ is given by

\begin{equation}
Q_{n}(q)\equiv 1.q.(2q-1).(3q-2)...[nq-(n-1)].
\end{equation}

J. Naudts [16-19] generalized the idea of q-deformed exponentials
and
logarithms to what is now called $\kappa -$deformed exponentials i.e., exp$%
_{\kappa }(x)$. A $\kappa -$deformed exponential is a convex
function with a value of one when its argument is zero. It is also
equal to or greater than zero for all real arguments. The $\kappa
-$deformed logarithm is defined in a similar manner. By choosing a
$\kappa -$deformed logarithm as

\begin{equation}
\ln _{\kappa }(x)=(1+\frac{1}{\kappa })(x^{\kappa }-1)\text{ \ \ \ \
\ \ \ \ \ \ \ \ \ \ \ \ \ \ \ \ \ \ \ \ \ \ -1
\mbox{$<$}%
}\kappa \text{%
\mbox{$<$}%
1,}
\end{equation}

and its inverse function as

\begin{equation}
\exp _{\kappa }(x)=[1+\frac{\kappa }{1+\kappa }x]_{+}^{1/\kappa
}\text{,\ \ \ \ \ \ \ \ \ \ \ }
\end{equation}

\bigskip where $[x]_{+}=\max \{0,x\}$ $,$ one immediately obtains the
nonextensive formalism by setting $\kappa $ equal to (q-1). For
$\kappa =0,$ they become the usual logarithmic and exponential
function respectively.

\bigskip Likewise, if one chooses for the deformed exponential

\begin{equation}
\exp _{\kappa }(x)=[\kappa x+\sqrt{1+\kappa ^{2}x^{2}}]^{1/\kappa
}\text{ \ \ \ \ \ -1
\mbox{$<$}%
}\kappa \text{%
\mbox{$<$}%
1 and }\kappa \neq 0,\text{\ \ \ \ \ \ \ \ \ }
\end{equation}
and the deformed logarithm as

\begin{equation}
\text{\ }\ln _{\kappa }(x)=\frac{1}{2\kappa }(x^{\kappa }-x^{-\kappa }),%
\text{ \ \ \ \ \ }
\end{equation}
one obtains Kaniadakis'deformed functions [20]. Again, in the limit
$\kappa =0$, these functions coincide with the usual logarithmic and
exponential functions. In the sense emphasized above, both of the
Tsallis and Kaniadakis formalisms can be explained within the scheme
of deformed functions approach proposed by Jan Naudts.

In Section II, we use quantum divergences in the nonextensive
formalism in order to calculate the purity of state in the case of
Werner states. We show that each of them provides us with a
different answer. This difference is resolved by considering the
first and third choices of internal energy constraint. Our final
observation is related to the fact that there is no nonextensive
relative entropy expression of Bregman type which is compatible with
third choice of constraint. We summarize the results in Section III.

\section{Relative Entropies, Fidelity and Constraints}

\bigskip One particularly important concept is the one of relative entropy:
In extensive statistics, the physical meaning of relative entropy is
the free energy difference. In addition to this, it plays a very
decisive role in the context of the second law of thermodynamics in
the nonextensive formalism. Indeed, Abe and Rajagopal showed that
there might be expected a violation of the second law if the
nonextensive index q is not in the range (0,2] [21]. To be able to
do this, they made use of the nonextensive relative entropy
definition of Csisz\'{a}r type [22]

\begin{equation}
\text{\ }I_{q}(\rho \parallel \sigma )=\frac{1}{q-1}\text{\
}[Tr(\rho ^{q}\sigma ^{1-q})-1].
\end{equation}

Recently, J. Naudts [18] and T. D. Frank [14, 23] gave an
alternative definition of relative entropy of Bregman type [24]
which reads

\begin{equation}
\text{\ }D_{q}(\rho \parallel \sigma )=\frac{1}{q-1}\text{\
}[Tr(\rho ^{q})-Tr(\rho \sigma ^{q-1})]-[Tr(\rho \sigma
^{q-1})-Tr(\sigma ^{q})].
\end{equation}

S. Abe [25] made use of Eq. (11) in order to calculate the purity of
Werner state in the nonextensive framework. For this purpose, he
used Werner state [26] which is given by the density matrix

\begin{equation}
\text{\ }\rho _{W}=F\mid \Psi ^{-}\rangle \langle \Psi ^{-}\mid +\frac{1-F}{3%
}(\mid \Psi ^{+}\rangle \langle \Psi ^{+}\mid +\mid \Phi ^{+}\rangle
\langle
\Phi ^{+}\mid +\mid \Phi ^{-}\rangle \langle \Phi ^{-}\mid ),\text{ \ \ }%
\frac{1}{4}\leq F\leq 1
\end{equation}

where

\begin{equation}
\mid \Psi ^{\pm }\rangle \equiv \frac{1}{\sqrt{2}}(\mid +-\rangle
\pm \mid -+\rangle ),
\end{equation}

and

\begin{equation}
\mid \Phi ^{\pm }\rangle \equiv \frac{1}{\sqrt{2}}(\mid ++\rangle
\pm \mid --\rangle ).
\end{equation}

\bigskip

F is the fidelity of $\rho _{W\text{ }}$with respect to the pure
reference state $\sigma =\mid \Psi ^{-}\rangle \langle \Psi ^{-}\mid
.$ For the time being, we restrict ourselves to the interval q$\in
(0,1)$ since Eqs. (11) and (12) should not be too sensitive to small
eigenvalues of the matrices as in Ref [25]. If we substitute Werner
states in order to find the degree of purity with respect to $\sigma
$ in Eq.(11), we obtain

\begin{equation}
\text{\ }I_{q}(\rho _{W}\parallel \mid \Psi ^{-}\rangle \langle \Psi
^{-}\mid )=\frac{1}{1-q}(1-F^{q}).
\end{equation}

Indeed, this is the result already obtained by Abe in Ref. [25]. Let
us consider the alternative definition of quantum divergence
proposed by Naudts and T. D. Frank in Eq. (12) and repeat the above
calculation by substituting $\sigma $, as defined earlier. First,
let us consider the first two terms on the right hand side of Eq.
(12)

\begin{equation}
Tr(\rho ^{q}-\rho \sigma ^{q-1})=\sum_{i}\langle i\mid (\rho
^{q}-\rho \sigma ^{q-1})\mid i\rangle ,
\end{equation}

\begin{equation}
\text{ \ \ \ \ \ \ \ \ \ \ \ \ \ \ \ \ \ \ }=Tr(\rho ^{q})-\langle
\Psi ^{-}\mid \rho \mid \Psi ^{-}\rangle .
\end{equation}

For the second term, we have

\begin{equation}
\text{ \ \ \ \ \ \ \ \ \ \ \ \ \ \ \ \ \ \ }Tr(\rho \sigma
^{q-1}-\sigma ^{q})=\sum_{i}\langle i\mid (\rho \sigma ^{q-1}-\sigma
^{q})\mid i\rangle =\langle \Psi ^{-}\mid \rho \mid \Psi ^{-}\rangle
-\langle \Psi ^{-}\mid \Psi ^{-}\rangle .
\end{equation}

\bigskip Summing all the terms above, we obtain

\begin{equation}
\text{\ }D_{q}(\rho \parallel \sigma )=\frac{1}{q-1}\text{\
}[Tr(\rho ^{q})-\langle \Psi ^{-}\mid \rho \mid \Psi ^{-}\rangle
]-\langle \Psi ^{-}\mid \rho \mid \Psi ^{-}\rangle +\langle \Psi
^{-}\mid \Psi ^{-}\rangle .
\end{equation}

\bigskip

Next, we make use of Werner states $\rho _{W}$ instead of the
generic $\rho $ in the expression above and obtain

\bigskip
\begin{equation}
\text{\ }D_{q}(\rho \parallel \sigma )=\frac{1}{q-1}\text{\ }[F^{q}+3(\frac{%
1-F}{3})^{q}-F]-(F-1).
\end{equation}

\bigskip

Obviously, Eq. (21) is different from Eq. (16). It leads to negative
values for q$\in (0,1)$ and F smaller than 1. This apparent
difference, at first glance, may look like a possible inconsistency
in the nonextensive formalism, but let us look closer at these two
different relative entropy expressions i.e., Eqs. (11) and (12) by
trying to solve them within a perturbative approach. The reason for
this is to ensure that all eigenvalues of $\sigma $ are different
than zero. From now on, we do not have to restrict ourselves to any
particular interval of q as long as it is not equal to 1. In order
to do this, let us rewrite $\sigma $ as

\begin{equation}
\sigma =(1-\epsilon )\mid \Psi ^{-}\rangle \langle \Psi ^{-}\mid +\frac{%
\epsilon }{3}(1-\mid \Psi ^{-}\rangle \langle \Psi ^{-}\mid ).
\end{equation}

This definition of  $\sigma $ corresponds to our earlier definition
when we set $\epsilon $ equal to zero. If we recalculate Eqs. (11)
and (12) now, we obtain

\begin{equation}
\text{\ }I_{q}(\rho _{W}\parallel \sigma )=\frac{1}{q-1}[(1-\epsilon
)^{q}F^{q}+\epsilon ^{1-q}(1-F)^{q}-1)],
\end{equation}

whereas Frank-Naudts version, we have

\begin{equation}
\text{\ }D_{q}(\rho _{W}\parallel \sigma )=\frac{1}{q-1}\text{\ [}%
F^{q}+3^{1-q}(1-F)^{q}-(1-\epsilon )^{q-1}F-(\frac{\epsilon }{3}%
)^{q-1}(1-F)]-(1-\epsilon )^{q-1}F-(\frac{\epsilon }{3})^{q-1}(1-F)+(1-%
\epsilon )^{q}+3^{1-q}\epsilon ^{q}.
\end{equation}

Now let us consider the limit $\epsilon \rightarrow 0$ for these two
distinct expressions of divergence assuming F to be between 0 and 1.
Then, for q$\in (0,1)$, we obtain

\begin{equation}
\text{\ }I_{q}(\rho _{W}\parallel \sigma )=\frac{1}{1-q}(1-F^{q}).
\end{equation}
and
\begin{equation}
\text{\ }D_{q}(\rho _{W}\parallel \sigma )=+\infty .
\end{equation}
On the other hand, for q values greater than 1, we have
\begin{equation}
\text{\ }I_{q}(\rho _{W}\parallel \sigma )=+\infty .
\end{equation}
and
\begin{equation}
\text{\ }D_{q}(\rho _{W}\parallel \sigma )=\frac{1}{q-1}\text{\ [}F^{q}+3(%
\frac{1-F}{3})^{q}-F]-(F-1).
\end{equation}

In order to understand the difference between Eqs. (11) and (12), we
consider first the maximization of the functional

\begin{equation}
\Phi ^{ordinary}=S_{q}-\alpha (\sum_{i}p_{i}-1)-\beta
(\sum_{i}p_{i}\varepsilon _{i}-U^{ordinary}),
\end{equation}

where the superscript ''ordinary'' indicates that we are using
ordinary expectation values. The equation above provides us the
following solution

\begin{equation}
\widetilde{p_{i}}^{ordinary}=[1+(1-q)\widetilde{S}_{q}^{ordinary}]^{1/(q-1)}%
\times \lbrack 1-\frac{q-1}{q}\beta \prime (\varepsilon _{i}-\widetilde{U}%
^{ordinary})]_{+}^{1/(q-1)},
\end{equation}

where tilda denotes that this particular expression is calculated in
terms
of the maximum entropy distribution $\widetilde{p_{i}}^{ordinary}$ , and $%
\beta \prime $ is given by

\begin{equation}
\beta \prime =\frac{\beta
}{\sum_{i}(\widetilde{p_{i}}^{ordinary})^{q}}.
\end{equation}

If the normalized q-expectation value [27] is employed, the
functional to be maximized will be of the form

\begin{equation}
\Phi ^{normalized}=S_{q}-\alpha (\sum_{i}p_{i}-1)-\beta (\frac{%
\sum_{i}p_{i}^{q}\varepsilon
_{i}}{\sum_{j}p_{j}^{q}}-U^{normalized}),
\end{equation}
which gives

\begin{equation}
\widetilde{p_{i}}^{normalized}=[1+(1-q)\widetilde{S}%
_{q}^{normalized}]^{1/(1-q)}\times \lbrack 1-(1-q)\beta ^{\ast
}(\varepsilon _{i}-\widetilde{U}^{normalized})]_{+}^{1/(1-q)},
\end{equation}

where

\begin{equation}
\beta ^{\ast }=\frac{\beta
}{\sum_{i}(\widetilde{p_{i}}^{normalized})^{q}}.
\end{equation}

If we now substitute $\widetilde{p_{i}}^{ordinary}$ given by Eq.
(23) into Eq. (12), we obtain\bigskip

\begin{equation}
D_{q}(p\parallel \widetilde{p_{i}}^{ordinary})=\beta (F_{q}^{ordinary}-%
\widetilde{F}_{q}^{ordinary}),
\end{equation}

i.e., the relative entropy of Bregman type is nothing but the
difference of free energies when the ordinary expectation value is
considered. This result has been derived also by T. D. Frank [28] by
using a different technique
than the one adopted in this paper. Similarly, if we now substitute $%
\widetilde{p_{i}}^{normalized}$ in Eq. (26) into Eq. (11), we get

\begin{equation}
I_{q}(p\parallel \widetilde{p_{i}}^{normalized})=\beta ^{\ast \ast
}(F_{q}^{normalized}-\widetilde{F}_{q}^{normalized}),
\end{equation}

where $\beta ^{\ast \ast }=\frac{\sum_{i}(p_{i})^{q}}{\sum_{i}(\widetilde{%
p_{i}}^{normalized})^{q}}\beta .$ Eq. (29) shows us that the
relative entropy of Csisz\'{a}r type is nothing but the difference
of free energies when the normalized q-expectation value is being
used. Therefore, as far as their physical meanings are concerned,
these two relative entropies are the same as Csisz\'{a}r type, being
connected with the normalized q-expectation values and Bregman type
is depending on ordinary expectation values (for details, see Ref.
[29]).

One can consider writing Eq. (28) in terms of normalized
q-expectation values as is suggested by J. Naudts: This is
tantamount to rewriting Eq. (22) by changing ordinary probability
distributions to escort probability distributions and replacing q by
1/q. The escort probability distribution, which will be obtained
after the maximization, must then be substituted into the following
quantum divergence of Bregman type

\begin{equation}
\text{\ }D_{q}(\rho \parallel \sigma )=\frac{1}{(1/q)-1}\text{\
}[Tr(\rho ^{1/q})-Tr(\rho \sigma ^{(1/q)-1})]-Tr(\rho \sigma
^{(1/q)-1})-Tr(\sigma ^{1/q}).
\end{equation}

If one now substitutes\bigskip\ the new escort probability
distribution
which is obtained by the substitutions $p_{i}\rightarrow \frac{p_{i}^{q}}{%
\sum_{j}p_{j}^{q}}=P_{i}$ and $q\rightarrow 1/q$ into the equation
above, we obtain the same free energy difference which has been
obtained in Eq. (28)
but this time with terms such as F$_{q}^{normalized}$ and $\widetilde{F}%
_{q}^{normalized}$ i.e., normalized free energy expressions. This
{\it prima facie} looks like the solution to our problem: We have
obtained a relative entropy expresion in accordance with the third
constraint, simply making some minor changes in the form of Eq. (12)
and in the maximization procedure as explained above. This
apparently simple solution has two flaws: First of all, in the last
maximization procedure, we have used $\sum_{i}P_{i}=1$. But, this
simply means that we have chosen it \ as a constraint in our
maximization. Now reminding ourselves that $\frac{p_{i}^{q}}{%
\sum_{j}p_{j}^{q}}=P_{i}$, we see that what we used as a constraint
is nothing but an identity. If we choose to use this normalization
as a constraint, then we must treat $P_{i}$ as an independent
variable. Once this had been assumed, the entropy written in terms
of P$_{i}^{\prime }s$ will be a new entropy which is different from
Tsallis entropy. Secondly, S$_{q}(P)$ occurs to be nonconcave in
some intervals which has been shown by Abe [30] and Di Sisto et al.
[31]. All these observations make it clear that we still do not have
a plausible relative entropy definition of Bregman type.

\section{RESULTS AND DISCUSSIONS}

We have studied two definitions of relative entropy in current use
in the nonextensive formalism. We have calculated the degree of
purification of Werner state using quantum divergence as recently
proposed by Jan Naudts and T. D. Frank and showed that it provides
us a result different from the one obtained by S. Abe in Ref. [25]
using the quantum divergence of Csisz\`{a}r type. This difference is
traced to the fact that one of Bregman type is written in a way
compatible with the first choice and the other is compatible with
the third choice of constraint in the nonextensive formalism. The
Bregman type gives rise to solutions for q values greater than 1 and
F smaller than 1 whereas Csisz\`{a}r type is suitable for q values
between 0 and 1. They differ in their qualitative behavior too
because the Bregman type decreases as fidelity F decreases while the
Csisz\`{a}r type increases as F decreases. This too is related to
the choice of constraints. The problem of writing a relative entropy
expression of Bregman type in the nonextensive formalism compatible
with the third constraint is still an open question for further
research.

\section{ACKNOWLEDGEMENTS}
This research was partially supported by the Scientific and
Technological Research Council of Turkey. We also thank Jan Naudts
for pointing us the perturbative calculation included in this paper
and many insightful comments.

\end{document}